\documentstyle[subeqn,preprint,aps,epsfig]{revtex}

\begin{document}
\newcommand{\be}{\begin{equation}}
\newcommand{\ee}{\end{equation}}
\newcommand{\bq}{\begin{eqnarray}}
\newcommand{\eq}{\end{eqnarray}}
\newcommand{\Dsl}{D\!\!\!\!/\,}
\newcommand{\p}{\varphi}
\draft
\tighten
\title{Gravitinos in non-Ricci-flat backgrounds}
\author{David Nolland\footnote{E-mail: \tt d.j.nolland@durham.ac.uk}}
\address{Department of Mathematical Sciences,\\ Durham University, 
South Road, Durham\\ DH1 3LE, U.K.}
\date{May 2000}
\maketitle 

\begin{abstract}
We discuss the gauge invariance and ``mass'' of the
Rarita-Schwinger field in a background spacetime which is assumed to be
Einstein but not necessarily Ricci-flat.
\end{abstract}

Many elementary discussions of the spin 3/2 Rarita-Schwinger field begin by
assuming that the spacetime background is Ricci-flat. This is a necessary
condition for the existence of a reducible, massless RS field in four
dimensions \cite{duff} (i.e. Ricci-flatness is forced
on us unless we use the harmonic gauge $\gamma\cdot\psi=0$). This special
case is of limited interest, however, not least because many interesting
supergravity backgrounds involve spaces of constant curvature, such as
spheres and/or (Anti-)de-Sitter space.

In this letter we wish to clarify the issue of gravitino mass in a more
general background. The Rarita-Schwinger field does not seem to couple
consistently to non-gravitational external fields, and in the absence of
self-interactions the
equations of motion entail consistency conditions \cite{duff} which require the
background to be Einstein; $R_{\mu\nu}=g_{\mu\nu}R/d$, where $d$ is the
dimension of spacetime. So this is the most general possibility at the
level of linearized supergravity, for example.

The most general form of the action contains two mass
terms, but we will find that one of these can be shown by a change of field
variables to be equivalent to a decoupled spin-half fermion. We show that
the action is gauge invariant when the remaining mass term takes a value
which is proportional to the square root of the Ricci scalar. For all mass
values the Lagrangian can be written in a Dirac-like form.

\medskip

With a Euclidean signature we start with the Lagrangian \cite{van}
\be
\label{lag}
{\cal L}=\sqrt g
(\bar\psi_\mu\gamma^{\mu\nu\rho}D_\nu\psi_\rho-m\bar\psi_\mu\gamma^{\mu\nu}\psi_\nu-\bar m\bar\psi^\mu\psi_\mu),
\ee
and make the decomposition
\be
\label{phi}
\psi_\mu=\p_\mu+{D^T_\mu\over D\cdot D^T}D^T\cdot\psi+{1\over d}\gamma_\mu\gamma\cdot\psi,
\ee
where $D^T_\mu=(\delta^\nu_\mu-\gamma_\mu\gamma^\nu/d)D_\nu$ is
$\gamma$-transverse so that 
\be
\label{const}
\gamma\cdot\phi=D\cdot\phi=0.
\ee 
Then (\ref{lag}) becomes
\bq
\label{newlag}
{\cal L}&=&\sqrt g\biggl[\bar\phi^\mu(\Dsl +m)\phi_\mu+\bar\psi\cdot D^T\biggl({{d-2\over
  d}\Dsl-m+\bar m\over D\cdot D^T}\biggr)D^T\cdot\psi-{d-2\over d}\bar\psi\cdot
D^T\gamma\cdot\psi\nonumber\\& &-{d-2\over d}\bar\psi\cdot\gamma
D^T\cdot\psi+\bar\psi\cdot\gamma\biggl({(d-1)(d-2)\over d^2}\Dsl+{d-1\over
  d}m+{1\over d}\bar m\biggr)\gamma\cdot\psi\biggr].
\eq
Now if we put $\psi_1=\sqrt{D\cdot D^T}D^T\cdot\psi$ and
$\psi_2=\gamma\cdot\psi$ then the change of variables $\psi_\mu\rightarrow
(\phi_\mu,\psi_1,\psi_2)$ has a trivial Jacobian, and the coupled pair
$(\psi_1,\psi_2)$ can be diagonalized from (\ref{newlag}), using $D\cdot
D^T={d-1\over d}\Dsl^2-{1\over4}R$. One of the
resulting fields is a trivial auxiliary field; the other has the Lagrangian
\be
{\cal L}=\sqrt g\bar\psi\biggl({d-2\over d}\bar m\Dsl-{d-1\over d}m^2+{d-2\over
  d}m\bar m+{1\over d}\bar m^2+{(d-2)^2\over 4d^2}R\biggr)\psi.
\ee
If $\bar m\ne 0$ then we can normalize $\psi$ and the original Lagrangian
(\ref{lag}) is therefore equivalent to
\be
{\cal L}=\sqrt g\bigl[\bar\phi^\mu(\Dsl+m)\phi_\mu+\bar\psi(\Dsl+M)\psi\bigr],\quad M={\bar
  m\over d-2}+m-{(d-1)m^2\over (d-2)\bar m}+{(d-2)\over4d\bar m}R.
\ee

So the effect of having $\bar m\ne 0$ in (\ref{lag}) is the same as adding an
additional spin-half fermion of mass $M$. Note that this conclusion is unaffected by the
introduction of interaction terms which would merely couple $\psi$ to
$\phi_\mu$. Thus we may without loss of generality set $\bar m=0$. 

Having done so, we conclude that $(\psi_1,\psi_2)$ produces {\em two\/}
auxiliary fields, unless $M\bar m=0$, in which case one of them decouples
completely from the action, signalling the presence of a gauge
invariance. Indeed, we can easily verify that for
\be
\label{massless}
m=\pm{1\over2}(d-2)\sqrt{R\over d(d-1)},
\ee
(\ref{lag}) is invariant under
\be
\delta\psi_\mu=D_\mu\lambda+{m\over d-2}\gamma_\mu\lambda.
\ee
Using the methods of \cite{nkn} it is straightforward to show that the
Lagrangian in the ``Feynman'' gauge \cite{endo} can still be written in the
Dirac form, with two Fadeev-Popov ghosts of mass $\quad
\pm{1\over2}\sqrt{dR/(d-1)}$ and a ``gauge-fixing'' commuting spinor ghost of
mass $5m/(d-2)$. In this case the field in the Lagrangian is unconstrained. Of
course there is nothing to stop us making another choice of gauge, in which
case the operator in the Lagrangian takes a slightly different form
\cite{endo2}.

For the massive gravitinos we still have the constraints (\ref{const}). By
introducing Lagrange multiplier fields for these quantities we find that we
can remove the constraints at the cost of introducing a pair of ghosts with
masses $\pm\sqrt{m^2+R/d}$. For the ``massless'' values (\ref{massless}) these
coincide with the Fadeev-Popov ghosts. So 
for {\em all\/} values of the mass we can write the Lagrangian
as\footnote{Kaluza-Klein compactifications of supergravity \cite{kim} tend to
yield gravitinos satisfying precisely the equations of motion implied by
(\ref{final}) and (\ref{const}).}
\be
\label{final}
{\cal L}=\sqrt g\bar\phi^\mu(\Dsl+m)\phi_\mu,
\ee
where $\phi$ satisfies (\ref{const}). If $m$ satisfies (\ref{massless}) we
have a single ghost of mass $5m/(d-2)$.

\medskip

In conclusion, we have shown that the most general form of the
Rarita-Schwinger Lagrangian with arbitrary dimension and mass can be written
in the Dirac form (\ref{final}). We identify the ``massless'' gravitino as
the one of mass $m={1\over2}(d-2)\sqrt{R/d(d-1)}$, for which there is a single ghost of mass
$5m/(d-2)$. There are also ghosts which are equivalent to imposing the constraints
(\ref{const}). 

All this immediately allows many results involving anomalies,
etc. \cite{endo,others} to be extended to the
case of non-Ricci-flat spacetimes. It is also relevant to the study of
supergravity compactifications
beyond tree-level, as in \cite{us}, which is especially interesting in the
context of the AdS/CFT correspondence \cite{Maldacena}.

\section*{Acknowledgement}

The author is indebted to Paul Mansfield for useful conversations.

\end{document}